\documentclass{aa}
\usepackage{txfonts}
\usepackage{inputenc}
\usepackage{natbib}
\bibpunct{(}{)}{;}{a}{}{,}
\usepackage{graphicx}

\begin{document}

\title{New constraints on dust grain size and distribution in CQ Tau}
\author{Andrea Banzatti\inst{1,2} \and Leonardo Testi\inst{2,3} \and Andrea Isella\inst{4} \and Antonella Natta\inst{3} \and Roberto Neri\inst{5} \and David J. Wilner\inst{6}}
\institute{ETH Zurich, Wolfgang-Pauli-Strasse 27, CH-8093 Zurich, Switzerland 
\and 
ESO, Karl Schwarzschild str. 2, D-85748 Garching bei Muenchen, Germany 
\and 
INAF - Osservatorio Astrofisico di Arcetri, Largo E. Fermi 5, I-50125 Firenze, Italy
\and 
Division of Physics, Mathematics and Astronomy, California Institute of Technology, MC 249-17, Pasadena, CA 91125, USA 
\and 
Institut de Radioastronomie Millim\'etrique, 300 rue de la Piscine, F-38406 Saint Mart{\'\i}n d'H\`eres, France 
\and 
Harvard-Smithsonian Center for Astrophysics, 60 Garden Street, MS 42, Cambridge, MA 02138, USA}
\offprints{ltesti@eso.org}
\date{Received ... / Accepted ...}
\abstract 
{Grain growth in circumstellar disks is expected to be the first step towards the formation of planetary systems.
There is now evidence for grain growth in several disks around young stars.}
{Radially resolved images of grain growth in circumstellar disks are believed to be a powerful tool to 
constrain the dust evolution models and the initial stage for the formation of planets. In this paper we attempt to 
provide these constraints for the disk surrounding the young star CQ~Tau. This system was already suggested from previous studies to host a population of grains grown to large sizes.}
{We present new high angular resolution ($0.\!\!^{\prime\prime}3-0.\!\!^{\prime\prime}9$) observations at wavelengths from 850~$\mu$m to 3.6~cm obtained at the SMA, IRAM-PdBI and NRAO-VLA interferometers. We perform a combined analysis of the spectral energy distribution and of the high-resolution images at different wavelengths using a model to describe the dust thermal emission from the circumstellar disk. We include a prescription for the gas emission from the inner regions of the system. } 
{We detect the presence of evolved dust by constraining the disk averaged dust opacity coefficient $\beta$ (computed between 1.3 and 7~mm) to be 0.6$\pm$0.1. This confirms the earlier suggestions that the disk contains dust grains grown to significant sizes and puts this on firmer grounds by tightly constraining the gas contamination to the observed fluxes at mm-cm wavelengths. We report some evidence of radial variations in dust properties, but current resolution and sensitivity are still too low for definitive results.}
{}
\maketitle


\section{Introduction}

In recent years the study of planetary system formation has grown enormously. 
While it is quite firm that the formation of planets should occur during the pre-main sequence (PMS) phase of the central star, when the circumstellar disk is still rich in gas and dust, our understanding of the complex variety of factors which contribute to that process is still very sketchy. Two possible scenarios are currently considered for planet formation: the \textit{``disk instability"} for gas giant planets \citep[see e.g.][]{2007prpl.conf..607D}, and the \textit{``core accretion"} for solid planetary cores from the dust particles in the disk \citep[see e.g.][]{1993ARA&A..31..129L}. 
The growth by coagulation of dust particles in the circumstellar matter from the typical sub-micron size found in the interstellar matter (ISM) to millimeter and centimeter sizes is now considered as the first step in the latter scenario; constraining observationally the dust evolution in the disk midplane allow us to put constraints
on the planet formation theories \citep[see][]{2000prpl.conf..533B}. 

Following the initial attempts to characterize the dust population using the submm spectral energy distribution of disks  \citep[e.g.][]{1991ApJ...381..250B,1993Icar..106...20M}, the last decade has seen a
number of studies dedicated to the search for large grains in protoplanetary disks at mm and cm wavelengths.
These studies have been made possible by the improved capabilities of millimeter and centimeter wave
arrays, but are still limited to bright objects \citep{2000ApJ...534L.101W, 2005ApJ...626L.109W,2001ApJ...554.1087T,2003A&A...403..323T,2006A&A...446..211R,2007prpl.conf..767N,2009A&A...495..869L,2010A&A...512A..15R}. Even in these favourable cases it has been so far impossible to resolve the radial variations of the dust properties and high angular resolution observations have mostly been used to constrain the mass distribution in disks by resolving the surface brightness profile at a single wavelength \citep{2007A&A...469..213I,2008ApJ...678.1119H,2009ApJ...700.1502A,2009ApJ...701..260I}. Resolving the radial variations of the dust properties is particularly interesting to constrain models of dust evolution in protoplanetary disks that include fragmentation and radial drift of large grains \citep[see e.g.][]{2008A&A...480..859B,Bea2010a}. Very recently \citet{Iea2010} used high sensitivity and high angular resolution CARMA observations at 1.3 and 3~mm to set limits on the possible radial variations of the dust properties in the protoplanetary disks around RY~Tau and DG~Tau.

In this paper we focus on the protoplanetary disks of CQ Tau in the Taurus-Auriga star-forming region.
CQ~Tau is one of the closest \citep[100~pc, Hipparchos,][]{1997A&A...323L..49P} young stellar objects with disk 
observable from the northern hemisphere, where most of the current high sensitivity and high spatial resolution
millimeter and centimeter wave arrays are located. The central star is a well known variable of the UX~Ori class with spectral type A8 and estimated age $\sim 10$~Myr. The star-disk system has been studied in detail by
various authors and the presence of grain growth to large sizes ($\sim$cm) was inferred from the mm 
range of the SED 
\citep[][]{2001ApJ...547.1077C,2000prpl.conf..559N,2001ApJ...554.1087T,2003A&A...403..323T,2008A&A...488..565C}. 
Here we refine  the investigation of dust 
properties in the disk of CQ Tau using new high-resolution observations obtained from the VLA, the PdB 
and the SMA interferometers at cm and mm wavelengths (Sections~\ref{sec:obs} and \ref{sec:res}). We 
combine our 
new observations with archival and literature ones and re-analyze them together. We use a 
dusty disk model with 
an accurate parametrization for the dust opacity. We include a prescription to account for
the gas emission. 
The average grain properties are first constrained by fitting the observed
SED (Section~\ref{sec:fit_sed}), separate model fits to the visibilities at various wavelengths
are then used to contrain the disk structure and to probe possible radial variations of 
dust properties (Section~\ref{sec:fit_vis}). 


\section{Observations and data reduction}\label{sec:obs}
\subsection{VLA observations}

The new observations at 1.3 cm (22.4 GHz, K-band) and 3.6 cm (8.4 GHz, X-band) were done on 
September 11 and 14, 2006 and November 03, 08 and 12, 2007 with the NRAO\footnote{The National Radio Astronomy Observatory is a facility of the National Science Foundation operated under cooperative agreement by Associated Universities, INC.} VLA in Soccorro. The array was in the B configuration, with a range of baselines from 210 m to 11.4 km providing a maximum angular resolution of 0.7 and 0.3 arcsec at 3.6 and 1.3 cm respectively. CQ Tau had been already observed in both bands with the VLA at lower angular
resolution, but only an upper limit at 3.6 cm was available from previous observations \citep[][]{2001ApJ...554.1087T}, while the previous 1.3~cm observations were not published. To maximize the signal-to-noise we decided to combine all VLA data (see Table \ref{table:calib_data}). The data reduction 
was performed using the CASA software package, following the standard procedures as described in the Data Analysis Cookbook\footnote{The Common Astronomy Software Applications (CASA) manual is available on \textit{casa.nrao.edu}}. At the time of this work CASA is still under development
so we compared step by step the reduction products using AIPS and with our work we confirm that the two packages yield to consistent results. Standard calibration procedures were employed, the complex gain calibrators used and their fluxes as derived from comparison with the VLA primary calibrators observations
are reported in Table~\ref{table:calib_data}, flux calibration is expected to be accurate within 5\%\ at 3.6~cm
and 10\%\ at 1.3~cm.  For deconvolution and imaging the natural weighting of visibilities was used in all cases. Once calibrated, all datasets belonging to the same band were added together for the final deconvolution and imaging, so that the final maps of CQ Tau at 1.3 and 3.6 cm result from the combination of all VLA data calibrated in CASA (see Fig.\ref{figure:maps}).


\subsection{PdBI observations}

Observations of CQ Tau were carried out simultaneously at 2.7~mm and
1.3~mm in the most extended configuration of the six-element IRAM
Plateau de Bure interferometer on Jan 15, 2006. Conditions during the
observations were excellent as testified by the high atmospheric
transparency (pwv$=0.5-1.0$~mm for 70\%\ of the time) and high
atmospheric phase stability ($\sigma_\phi \le 40^\circ$ at 1.3~mm)
observed over the 7.5~hours track. The spectral
configuration was set to cover a contiguous band of 550~MHz for
maximum continuum sensitivity. The RF calibration was measured on
3C454.3, and amplitude and phase calibration were made on
0528+134. The absolute flux calibration was determined using as
primary calibrator MWC349. Data reduction and calibration were made
using the GILDAS software package in the standard antenna based
mode. Continuum point source sensitivities of 0.15~mJy/beam and 
0.5~mJy/beam were obtained at 2.7~mm and 1.3~mm, respectively, and 
consistent with the measured system temperatures (250-300~K).

\subsection{SMA observations}

The CQ Tau system was observed with the Submillimeter Array (SMA) on Mauna
Kea, Hawaii, on 2008 Jan 24. The array was in the extended configuration, and
the six available antennas provided baseline lengths from 60 to 225 meters.
The atmospheric transparency was excellent, with 225~GHz opacity of 0.05 as
measured at the nearby Caltech Submillimeter Observatory. The double sideband
receivers were tuned with a local oscillator (LO) frequency of 340.8~GHz,
and the correlator provided 2~GHz of bandwidth in sidebands $\pm5$~GHz from
the LO frequency. Useable data were obtained for CQ~Tau over an hour angle
range of $-0.9$ to $+2.2$. Complex gain calibration used observations of the
calibrators J0530+135 (1.75~Jy) and 3C111 (3.0~Jy) interleaved with CQ~Tau.
The passband response was calibrated using observations of the strong source
3C273. The flux scale was determined by bootstrapping observations of 3C273
calibrated against planets and is accurate at the 15\%\ level. All of the data
editing and calibration were performed using standard routines in the MIR
software package for IDL.

\section{Results}\label{sec:res}

\begin{figure} 
\centering
\resizebox{1.0\hsize}{!}{\includegraphics[angle=270]{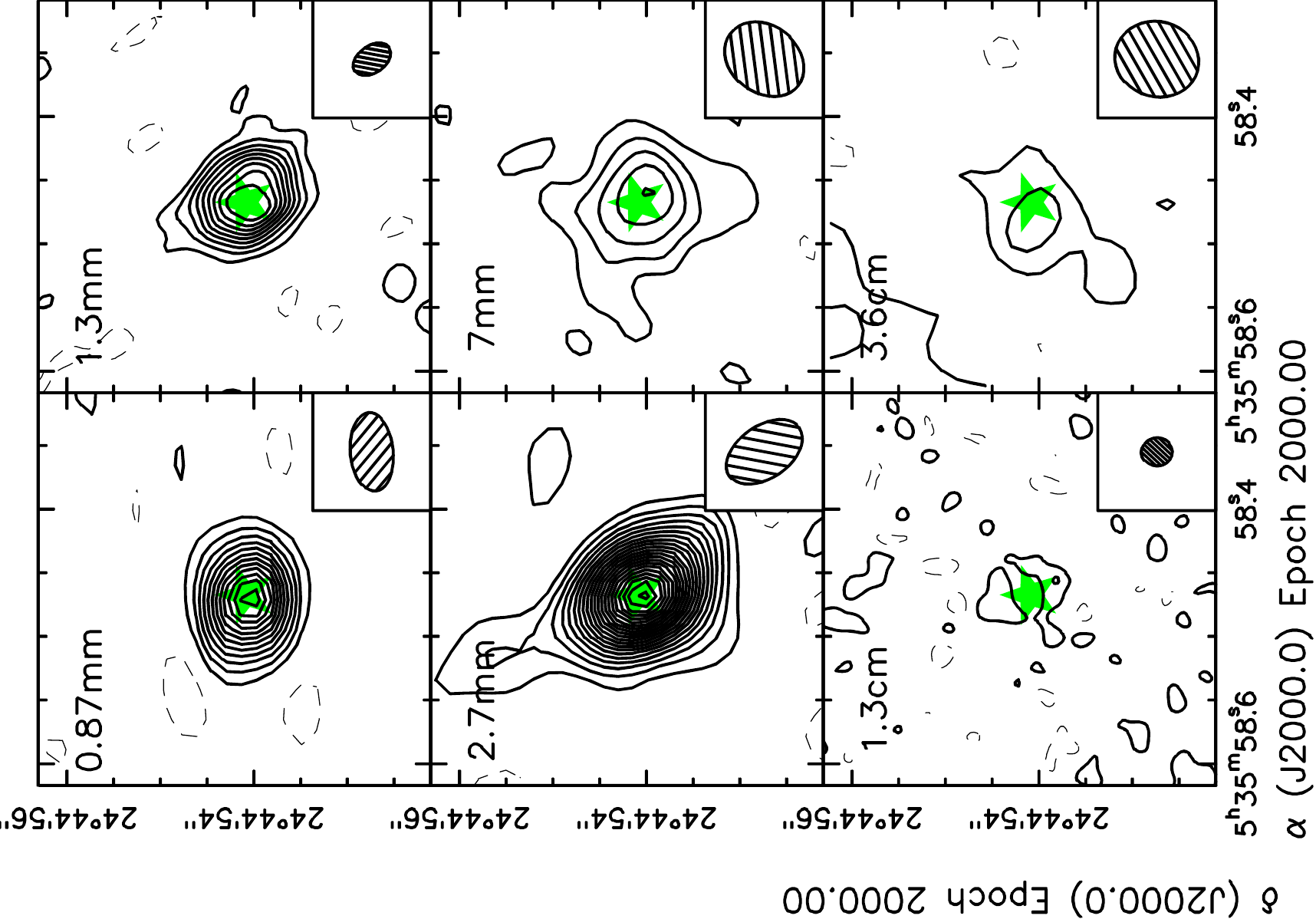}} 
\caption{CQ Tau continuum maps in mm-cm bands from our new high angular resolution data (from top left to bottom right): 0.87 mm, 1.3 mm, 2.7 mm, 7 mm \citep[from][]{2003A&A...403..323T}, 1.3 cm, 3.6 cm maps. In each plot the position of the central star and the beam ellipse are shown. Contours levels are spaced by 2$\sigma$ where $\sigma$ is 9 mJy at 0.87 mm, 1 mJy at 1.3 mm, 0.3 mJy at 2.7 mm, 0.1 mJy at 7 mm, 0.02 mJy at 1.3 cm and 0.009 mJy at 3.6 cm. All maps have been corrected from their original epoch to the epoch 2000.0, using the proper motion of the optical star as measured by \textit{Hipparcos} (RA: 2.67, Dec: -26.17 mas/year), the star is also plotted at the epoch 2000.0 coordinates, as given by Simbad at CDS.} 
\label{figure:maps} 
\end{figure} 

In Figure~\ref{figure:maps} we show the interferometric images of the CQ~Tau system from 
0.87~mm through 3.6~cm. We detect strong emission at millimeter wavelengths and the emission
is resolved at all wavelengths from 0.87~mm through 7~mm \citep[as discussed in][]{2003A&A...403..323T}.

At 1.3 mm, we measure a spatially integrated flux of 103 mJy, which is $\sim$40\% smaller than that measured 
at 2\arcsec resolution with the same instrument \citep[IRAM-PdBI,][]{2008A&A...488..565C}. 
Even if this discrepancy may be marginally within the calibration uncertainties at this wavelength (10-20\%), we
caution that the difference may be caused by the presence of extended ($\ge 3\arcsec$) dust emission 
partially filtered by our observations. This however does not seem to be consistent with measurements at the 
other wavelengths (which do not show extended emission) and our model fit to the SED 
(see Sect.~\ref{sec:fit_sed}) which does not seem to require additional extended components.

At centimeter wavelengths, our new VLA observations allowed us to detect the system for the first time 
at wavelengths longer than 7~mm, with a detection at both 1.3 and 3.6~cm. 
The source is not resolved at 3.6~cm; at 1.3~cm we tentatively detect extended emission
at a $\sim$2$\sigma$ level around the optical star position. In Table~\ref{table:data_summary} we report the beam sizes and
sensitivities of our maps as well as peak and integrated fluxes for CQ~Tau, in the same table we also report the
fluxes from the literature that we have used in Sect.~\ref{sec:fit_sed} for the fits to the SED. 

The possible detection of an extended ring-like emission at 1.3~cm coincident with the disk detected at shorter wavelengths, if confirmed, suggests that at this wavelength we are still detecting the dust thermal emission.
The 3.6~cm emission is unresolved and it is consistent with emission from ionized gas in the inner region of 
a wind or around the central star. 

After adjusting the astrometric calibration of the radio observations to account for the measured proper
motion of CQ~Tau from Hipparchos \citep{1997A&A...323L..49P}, the optical position of the star is 
coincident with the peak emission at (sub-)millimeter wavelengths. 

\begin{table*}
\begin{minipage}[t]{2\columnwidth}
\caption{Summary of PdBI and VLA data (new + archival) reduced for this work. For each dataset we show the observation date, the array configuration, the primary flux calibrator used and the flux assumed from the observatory calibration plan, the amplitude and phase calibrator used and its derived flux at the time of the observations (note that we reported the calibrator names as in the databases of the observatories at the time of the observations, 0528+134 used at PdBI in 2006 is the same object as 0530+135 used at the VLA in 1999).}
\label{table:calib_data}
\centering
\renewcommand{\footnoterule}{}
\begin{tabular}{l l c c c c c c}  
\hline\hline
 Band & Project & Date & array  & \multicolumn{2}{c}{Flux calibrator} & \multicolumn{2}{c}{Gain calibrator} \\
   &  &  & conf. & Name & Flux (Jy) & Name & Flux (Jy) \\
\hline
1.3 mm & P054 & 06 Jan 15 & A & MWC349 & 1.9 & 0528+134 &2.1 \\ 
2.7 mm & P054 & 06 Jan 15 & A & MWC349 & 1.2 & 0528+134 &3.1 \\  
1.3 cm & AK0475 & 99 Jan 14 & C & 0137+331 & 1.117 & 0530+135 &2.77 \\ 
 & AT0327 & 06 Sep 14 & B & 0137+331 & 1.117 & 05595+23539 & 0.292\\
 & AT0352 & 07 Nov 03 & B & 0137+331 & 1.117 & 05595+23538 & 0.246\\
3.6 cm & AT0327 & 06 Sep 11 & B & 0137+331 & 3.164 & 05595+23539 & 0.463 \\
 & AT0352 & 07 Nov 08 & B & 0137+331 & 3.164 & 0547+273 & 0.242\\
 & AT0352 & 07 Nov 12 & B & -- & -- & 0547+273 & 0.242\footnote{In this dataset the flux calibrator is absent; for calibration we assumed the flux value of the phase-calibrator derived in the same band four days before (07 Nov 08).}  \\
\hline \hline
\end{tabular} 
\end{minipage}
\end{table*}

\begin{table}
\begin{minipage}[t]{1\columnwidth}
\caption{Summary of available data at mm--cm wavelengths; the new observations presented in this work are highlighted.}
\label{table:data_summary}
\centering
\renewcommand{\footnoterule}{}
\begin{tabular}{c c c c c cc}   
\hline\hline                         
  $\lambda$  & Synthesized beam \footnote{The first two values correspond to the size of the axes of the synthesized beam ellipse, while the third to its position angle. The beam is reported only for interferometric observations}  & Peak\footnote{Only for interferometric observations}  & Flux   & rms\footnote{In the case of data reduced in this work the $rms$ is extracted directly from the residual noise in final cleaned image, otherwise it is the uncertainty reported in literature.}  & Ref.\footnote{Data from the literature: (1): this work, (2): \citet{1997ApJ...490..792M}, (3): \citet{2008A&A...488..565C}, (4): \citet{2000ApJ...529..391M}, (5): \citet{2003A&A...403..323T}, (6): \citet{2001ApJ...554.1087T}.}\\
  (mm) &  (arcsec, deg ) & (mJy) & (mJy) & (mJy) &  \\
\hline
  0.87  &0.85 $\times$ 0.45, 95 & 253 & 421 & 9 & \textbf{(1)} \\
  1.3  &  ... & ... & 221 & 40 & (2) \\
  1.3  &   1.5  & ... & 162 & 2 & (3) \\
  1.3  &   2 & ... & 143 & 8.4 & (4) \\
  1.3  &   0.45 $\times$ 0.29, 35 & 18 & 103 & 1 & \textbf{(1)} \\
  2.7  &   0.91 $\times$ 0.56, 34 & 9 & 22 & 0.3 & \textbf{(1)} \\
  2.7  &  ... & ... & 18 & 0.6 & (2) \\
  3.4  &   ... & ... & 13.1 & 0.5 & (3) \\
  6.9  &   0.92 $\times$ 0.72, -37 & 1.0 & 2.0 & 0.1 &  (5) \\
  13.4  &   0.33 $\times$ 0.30, 2 & 0.08 & 0.28 & 0.02 & \textbf{(1)} \\
  35.7  &   0.91 $\times$ 0.81, -15 & 0.046 & 0.079 & 0.009 & \textbf{(1)} \\
  62.5  &   31 $\times$ 23, 49 & $<$ 0.75 & ... & 0.25 & (6) \\
\hline \hline
\end{tabular} 
\end{minipage}
\end{table}

\section{Dusty disk model, gas emission and SED}\label{sec:mod}
\subsection{Dusty disk}

\begin{figure*} 
\centering
\resizebox{1\hsize}{!}{\includegraphics{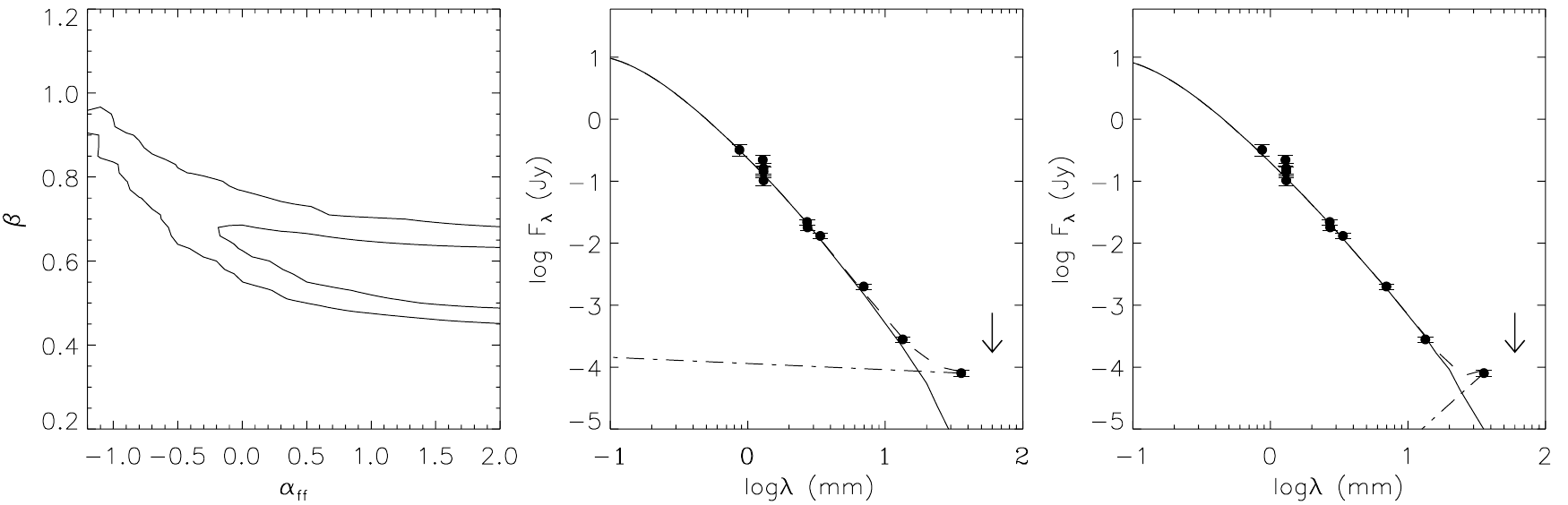}} 
\caption{Results of the SED fitting using the reference dust described in the text (Section \ref{sec:mod}). Left: confidence limits on dust opacity index $\beta$ and gas emission index $\alpha_{\rm{ff}}$; contours represent the 68\%\ and 95\%\ confidence levels for the fit (i.e. they represent the $\Delta\chi^2=2.3$, and 6.17 values). Middle and right: examples of fits to the SED within the 68\% confidence level limits; the dusty disk model is displayed as a continuum line, the free-free power law for the gas emission as a dotted-dashed line, and the total model (dusty disk + free-free) as a dashed line, the flux upper limit at 6 cm as an arrow. The middle and right panels are for $\alpha_{\rm{ff}}=-0.2$ and 2.0, respectively.} 
\label{figure:ref_dust} 
\end{figure*} 

To analyze the data we have used the model for passive flared two-layer disks presented by \citet{2001ApJ...560..957D}, based on \citet{1997ApJ...490..368C}. The model requires as input a set of parameters for the central star and the disk, which are used to compute the disk structure, SED and radial intensity profile at each wavelength. The stellar properties are assumed known as in \citet{2003A&A...403..323T} (T$_{\rm{eff}}$=6900 K, $L_{\star}$=6.6 L$_{\odot}$, $M_{\star}$=1.5 M$_{\odot}$, D=100 pc). The disk surface density distribution is assumed as $\Sigma (R)=\Sigma_{\rm 40AU} \left( R / 40 \rm{AU} \right) ^{-p}$, where $\Sigma_{\rm 40AU}$ is the surface density at the fiducial radius of 40 AU.
For the dust opacity at a given frequency $k_{\nu}$ we consider the parameterization by \citet{1993Icar..106...20M} as: $k_{\nu}=(\int n(a)a^{3}k_{\nu,1}(a)da)/(\int n(a)a^{3}da)$, where $a$ is the particle size distributed between a minimum and a maximum value $a_{min}$ and $a_{max}$ according to $n(a)\propto a^{-q}$, and $k_{\nu,1}$ is the opacity per unit mass of the single particle, which depends on grain shape and chemical composition. The value of $a_{min}$ is set to 0.01 $\mu$m according to the typical value observed for unprocessed dust in the ISM, while for the power law exponent we have used $q=3$ 
as proposed in previous studies and theoretical expectations \citep[see][]{2006ApJ...636.1114D}. 
In this work we assume as reference the same ensemble of dust used in \citet{2010A&A...512A..15R}, where porous spherical grains are composed by (volume percentage) 7$\%$ silicates, 21$\%$ carbonaceous materials, 42$\%$ water ice and 30$\%$ vacuum \citep[][]{1994ApJ...421..615P}. 
It is well known that the dust opacity, and thus the exact value of $a_{max}$ derived from fitting the millimeter dust emission, depends strongly on the dust composition and the power law index of the dust size distribution \citep[e.g.][]{2007prpl.conf..767N,2010A&A...512A..15R}. In Sect.~\ref{sec:fit_sed} we will show which is the range of uncertainty in our derivation of $a_{max}$ by exploring its dependence on the assumed dust model and the grain chemical/physical properties: removing the water ice, varying the vacuum volume from 10$\%$ to 50$\%$ and the grain size distribution index $q$ from 2.5 to 3.5. 
We stress that the observations constrain very well the dust opacity index $\beta$, in the simplified
first-order approximation $k_{\nu} \propto \nu^{\;\beta}$.
In this work we compute the slope of the dust opacity $\beta$ from the ratio of the values of $k_{\nu}$ at 1.3 and 7 mm. This provides a parameter that can be directly compared to other studies, but note that the shape of $k_{\nu}$ that we use in the disk model is the parameterization by \citet{1993Icar..106...20M} written above. The dependence of $\beta$ on $a_{max}$ and $q$ was discussed in \citet{2004ASPC..323..279N}, and  
\citet{2010A&A...512A..15R} showed the case of our reference ensemble of dust. 

\subsection{Gas emission}
Young stellar objects are known to show radio continuum emission from circumstellar gas,
this emission can either be thermal or non-thermal 
\citep[see e.g.][]{1987AJ.....93.1182A,1993ApJS...87..217S}. The radio emission can
normally be well approximated by a power law of the form $F_{\lambda}\propto \lambda^{\alpha_{\rm{ff}}}$
with negative values of the radio spectral index ($\alpha_{\rm{ff}}$) normally attributed to thermal emission \citep{1967ApJ...147..471M,1975A&A....39....1P,2000ApJ...542L.143F} and positive values to non-thermal emission \citep{1989A&A...217L...9G}.
\citet{1993ApJS...87..217S} found spectral indices spanning the range between -1.2 and 1.2 in a sample of Herbig Ae/Be stars. Using an equivalent parametrization \citet{2006A&A...446..211R} applied a simplified correction for free-free contamination at radio wavelengths (assuming a priori a 50\% free-free contribution at 1.3 cm), and found values for $\alpha_{\rm{ff}}$ between -0.1 and 0.4 in a sample of low-mass PMS stars in the Taurus-Auriga star forming region. 
These results show the variety of spectral indices displayed by radio emission in young stellar objects and
highlight the need of a proper characterization of the gas emission to accurately constrain the 
dust properties from millimeter and centimeter wave observations \citep[see e.g. the discussions by ][]{2001ApJ...554.1087T,2005ApJ...626L.109W}.
To properly estimate the contribution of the gas emission to the long wavelength SED of CQ~Tau, in our analysis 
we add to the dusty disk model a power law with variable spectral index $\alpha_{\rm{ff}}$ as above, which we 
try to constrain using our new observations at long wavelengths with the assumption that all the emission 
we measure at 3.6~cm is due to gas emission.
What we find with our analysis (which is presented in the next Section and in figures \ref{figure:ref_dust} and \ref{figure:var_dust}) is only an upper limit to the possible gas contribution at 7 mm and 1.3 cm.

\subsection{Model fitting to the SED}\label{sec:fit_sed}

We used the dusty disk model and the parametrization for the gas emission discussed above
to fit the submillimeter to centimeter wave SED of CQ~Tau. In Table~\ref{table:data_summary} we
show the flux used at the various wavelengths from our own data and from the literature.
The aim of SED fitting at mm-cm wavelengths is to constrain the maximum size of dust grains 
for the dust model we adopted (assuming a uniform dust population in the disk mid plane). 
To avoid degeneracies with the disk geometry, we assumed a disk outer radius of  63 AU,  a disk inclination of
inclination of 30$^\circ$, and a surface density distribution index $p = 0 $, as derived from the fit of the spatially 
resolved 1.3 mm dust emission as described in Section \ref{sec:fit_vis}. The inner radius is set to 0.2 AU, 
which corresponds to the dust evaporation radius as calculated in \citet{2006A&A...451..951I}. The disk
atmosphere is considered to be populated only with small dust grains with the same chemical composition 
as the midplane dust.
The SED is then fitted minimizing the $\chi^2$ as a function of the free parameters ($a_{max}$, $\Sigma_{40}$, 
and $\alpha_{\rm ff}$). 


The left panel of Figure~\ref{figure:ref_dust} shows the $\chi^2$ contours for
$\alpha_{\rm ff}$ and  $\beta$ assuming our reference dust. The data are consistent at the 68\%\ confidence level with 
$\alpha_{\rm ff}$ larger than $-0.1$ and $\beta$ between 0.49 and 0.68. 
For $\alpha_{\rm ff} > 1$ the free-free contribution at wavelengths up to 1.3 cm is 
negligible (less than $\sim$10\% at 1.3 cm) and $\beta$ becomes independent of $\alpha_{\rm ff}$. A gas emission index $\alpha_{\rm{ff}}$ higher than $-0.1$ implies 
that the free-free contamination to the observed flux is at most 5\% at 7 mm and 31\% at 1.3 cm. Such a low value of the free-free contribution at 1.3 cm would be consistent with our non-detection of an unresolved emission due to the gas at at this wavelength (as shown in Section \ref{sec:obs}), but, given the 
low signal to noise ratio of our 1.3~cm maps, future observations with higher sensitivity are needed to confirm this point.
Our constraints on $\beta$ are consistent and put on firmer grounds the previous estimates, which were 0.6$\pm0.1$ from \citet{2003A&A...403..323T} and 0.70$\pm0.04$ from \citet[note that these latter authors compute 
the value of $\beta$ between 1.3 and 3~mm]{2008A&A...488..565C}. Both of these previous studies could not constrain the cm emission and the free-free contamination at mm wavelengths. 

\begin{figure} 
\centering
\resizebox{1\hsize}{!}{\includegraphics{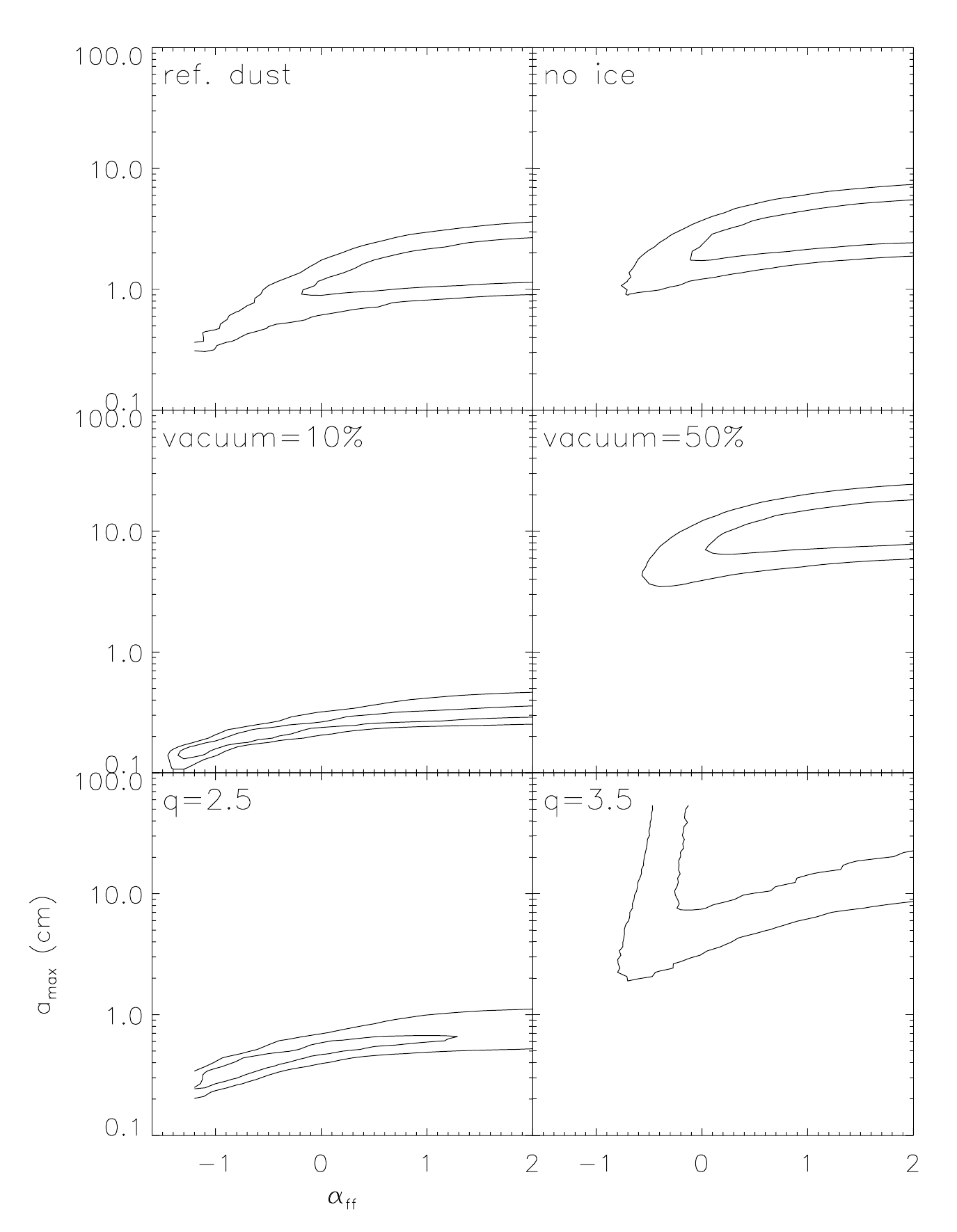}} 
\caption{Confidence limits on $a_{max}$ and $\alpha_{\rm{ff}}$ from the SED fitting using the different ensembles of dust; from top left to bottom right: reference dust with and without ice, volume percentage of vacuum 10\% and 50\%, grain size distribution index $q$=2.5 and $q$=3.5. Contours represent the 68\%\ and 95\%\ confidence levels for the fit.} 
\label{figure:var_dust} 
\end{figure}

As discussed by many authors \citep[see e.g. the discussion in ][]{2007prpl.conf..767N}, while the constrain on the value of $\beta$ is very strong from the millimeter SED fits, relating this to the exact properties of the 
dust grains in the disk is more difficult, due to the uncertainties in the dust composition (and hence opacity).
Fig.~\ref{figure:var_dust} top left panel shows the constraint on the maximum grain sizes using our reference
dust model, in this case we find that we can constrain the dust to be grown to $\sim$1~cm pebbles.
To explore how this conclusion depends on the dust properties we have also run a set of models varying the 
ice content, the level of porosity and the dust size distributions. 
In Fig.~\ref{figure:var_dust} we show the $\chi^2$ surfaces obtained exploring the constraints on $a_{max}$ and $\alpha_{\rm{ff}}$ for a representative set of dust parameters.
A summary of the results of all the fits is reported in Table~\ref{table:best_fit_sed}.
As expected, the grain sizes that we derive are larger in the case of more porous grains 
(no-ice and 50\%\ vacuum cases) and are smaller for more compacted grains (10\%\ vacuum case).
Derived grain sizes are also smaller if we assume that the grain size distribution is more heavily weighted 
towards large grains (q$=$2.5 case); while for a size distribution with q$=3.5$ very large grains, if present, 
do not contribute significantly to the dust opacity and we can only put a lower limit to the maximum grain 
size.
The result of this analysis is that, even considering variations to the properties of the dust, our data is 
only consistent with maximum grain sizes larger than about 1~mm, implying that the grain growth process
in this system is well advanced. In the remainder of this paper we will only consider the reference dust.

\begin{table*}
\begin{minipage}[t]{2\columnwidth}
\caption{Constraints on model parameters from the SED fitting. The range of values reported for each parameter is within the 68\% confidence level.}
\label{table:best_fit_sed}
\begin{tabular}{l|c c c c c c}
\hline
Parameter & ref. dust & no ice & vacuum 10\% & vacuum 50\% & $q$=2.5 & $q$=3.5 \\
\hline
$a_{max}$ (cm) & 0.8--2.4 & 1.7--4.7 & 0.1--0.3 & 6.5--16 & 0.2--0.7 & $\geq 7$ \\
$\alpha_{\rm{ff}}$ & $\geq -0.1$ & $\geq 0.0$ & $\geq -1.3$ & $\geq 0.1$ & -1.1--1.2 & $\geq -0.2$ \\
$\Sigma_{\rm 40}$ (g\,cm$^{-2}$)& 2.6--5.2 & 3.6--6.8 & 0.55--1.05 & 12.8--23.2 & 1.3--2.7 & $\geq 4.7$ \\
\hline
f-f$_{7mm}$ (\%) & $\lesssim$ 5  & $\lesssim$ 4 & $\lesssim$ 33 & $\lesssim$ 3 & 1--24 & $\lesssim$ 5 \\
f-f$_{1.3cm}$ (\%) & $\lesssim$ 31  & $\lesssim$ 28 & $\lesssim$ 100 & $\lesssim$ 26 & 9--84 & $\lesssim$ 35 \\
\hline
\end{tabular}
\end{minipage}
\end{table*}

\section{Model fitting to the visibility data}
\label{sec:fit_vis}

In the previous section we have shown that
the grains on the CQ Tau disk midplane have reached maximum sizes of at least a few mm, 
confirming earlier suggestions. This result was obtained by assuming values of the disk outer radius, inclination and surface density index
of 63 AU, 30$^{\circ}$ and 0, that are obtained as discussed below.

Following the procedure used in \citet{2003A&A...403..323T}, we produced model images corresponding to 
different sets of parameters: disk inclination and position angle (which we define in this paper as
the angle east of north of the line defining the intersection between the disk
plane and the plane of the sky),
disk radius, surface density power law index $p$ and normalization $\Sigma_{40}$.
For the dust properties we used the reference dust model discussed in the previous section 
for the disk mid plane and surface, assuming a constant composition along the entire disk. 
On the mid plane we adopted a maximum grain size of a$_{max}=$1~cm (as derived above).
The model images of the disks at each wavelength were Fourier transformed and sampled at the appropriate
positions on the (u,v) plane(s) corresponding to the observed samples. Following the method 
described in \citet{2007A&A...469..213I}, at each wavelength we computed
the $\chi^{2}$ function as:
\begin{eqnarray*}
 \chi ^{2}&=&\Sigma_{j} ( \left[  Re_{j}(obs) - Re_{j}(mod)\right] ^{2}\\
                       & &+\left[ Im_{j}(obs) - Im_{j}(mod)\right]^{2} ) /\sigma_{j}^{2}, 
\end{eqnarray*}
where $Re$ and $Im$ are the real and imaginary part of the observed ($obs$) and theoretical ($mod$)
correlated flux, and $\sigma$ is the noise on each measure. 

This procedure was repeated on a wide grid of model parameters to construct a $\chi^2$ hypercube.
In order to  find the best fitting model, the cube is then searched for the minimum of the $\chi^2$ as a 
function of all parameters. The procedure is repeated independently for all four datasets from 0.87~mm
to 7~mm and the best fitting model is derived independently at each wavelength. This procedure
has the advantage of allowing to search for systematic variations with wavelength of the model 
parameters. The drawback is that at each wavelength we are fitting a large set of parameters and 
some of these are only weakly constrained by some of our datasets.
The 1.3~cm dataset has a signal to noise ratio too low to perform a proper visibility 
fitting, while at 3.6~cm the emission is not due to the dust thermal emission in the disk, we thus
excluded the two longest wavelength datasets from the visibility fitting procedure. 

\begin{figure}
\includegraphics[angle=0, width=\columnwidth]{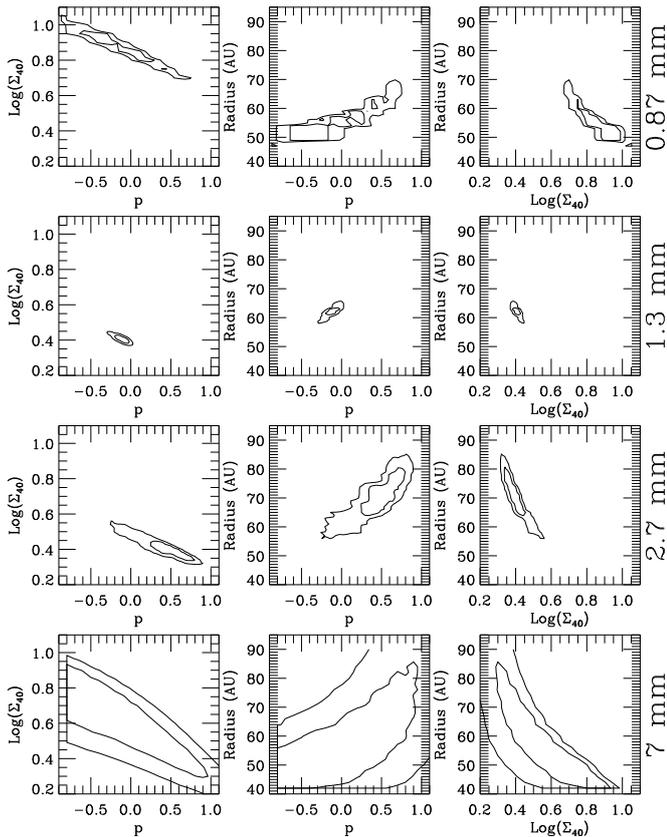}
\caption{$\chi^2$ hypercube projections on the (p,Log$_{10}(\Sigma_{40})$), 
(p,R$_{\rm out}$), and (Log$_{10}(\Sigma_{40})$,R$_{\rm out}$) planes, to produce each
plot we have chosen the value of the other parameters corresponding
to the minimum of $\chi^2$ at each position on the planes shown. Contours represent the 
68\%\ and 95\%\ confidence levels for the fit. 
In each row we show the results for the fit at a given wavelength, from top to bottom
ar 0.87~mm, 1.3~mm, 2.7~mm and 7~mm (as labelled to the right).} 
\label{fchi2} 
\end{figure}

\begin{figure} 
\resizebox{1.0\hsize}{!}{\includegraphics[angle=270]{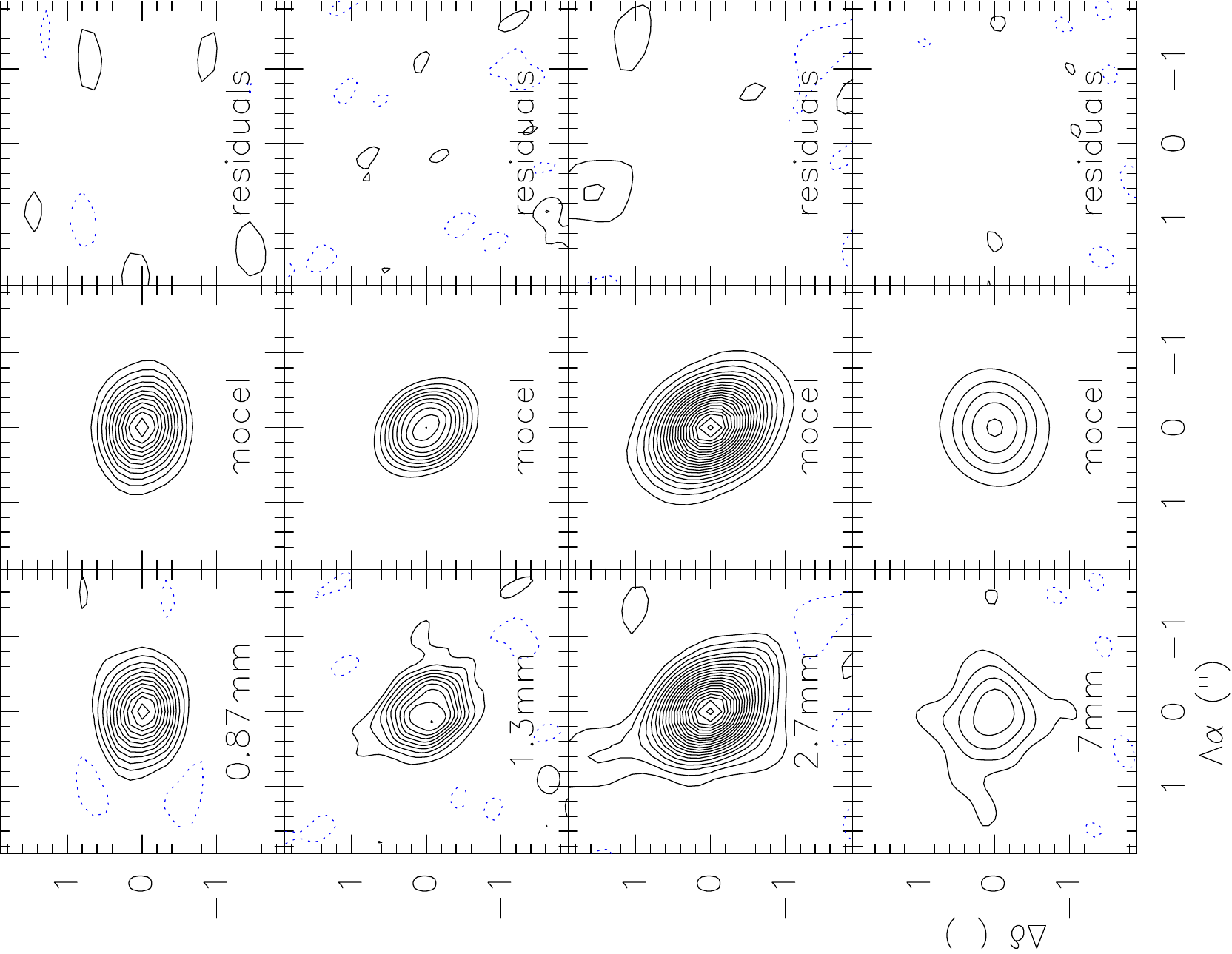}} 
\caption{Comparison between the observed images at 0.87~mm, 1.3~mm, 2.7~mm and 7~mm (left panels, from top 
to bottom) and the images corresponding to the best fit model at each wavelength obtained by sampling the 
model visibilities at the same locations and with the same noise as for the observed data (middle panels). The images
derived by the difference from the observed and model visibilities are shown on the right panels. At each 
wavelength the contours are the same in all three plots and correspond to $-2\sigma$ and from $2\sigma$ to the peak value spaced by $2\sigma$, where $\sigma$ is the rms measured in the observed map (see also Fig.~\ref{figure:maps}).} 
\label{fmodima} 
\end{figure}

\begin{figure} 
\resizebox{1.0\hsize}{!}{\includegraphics[angle=0]{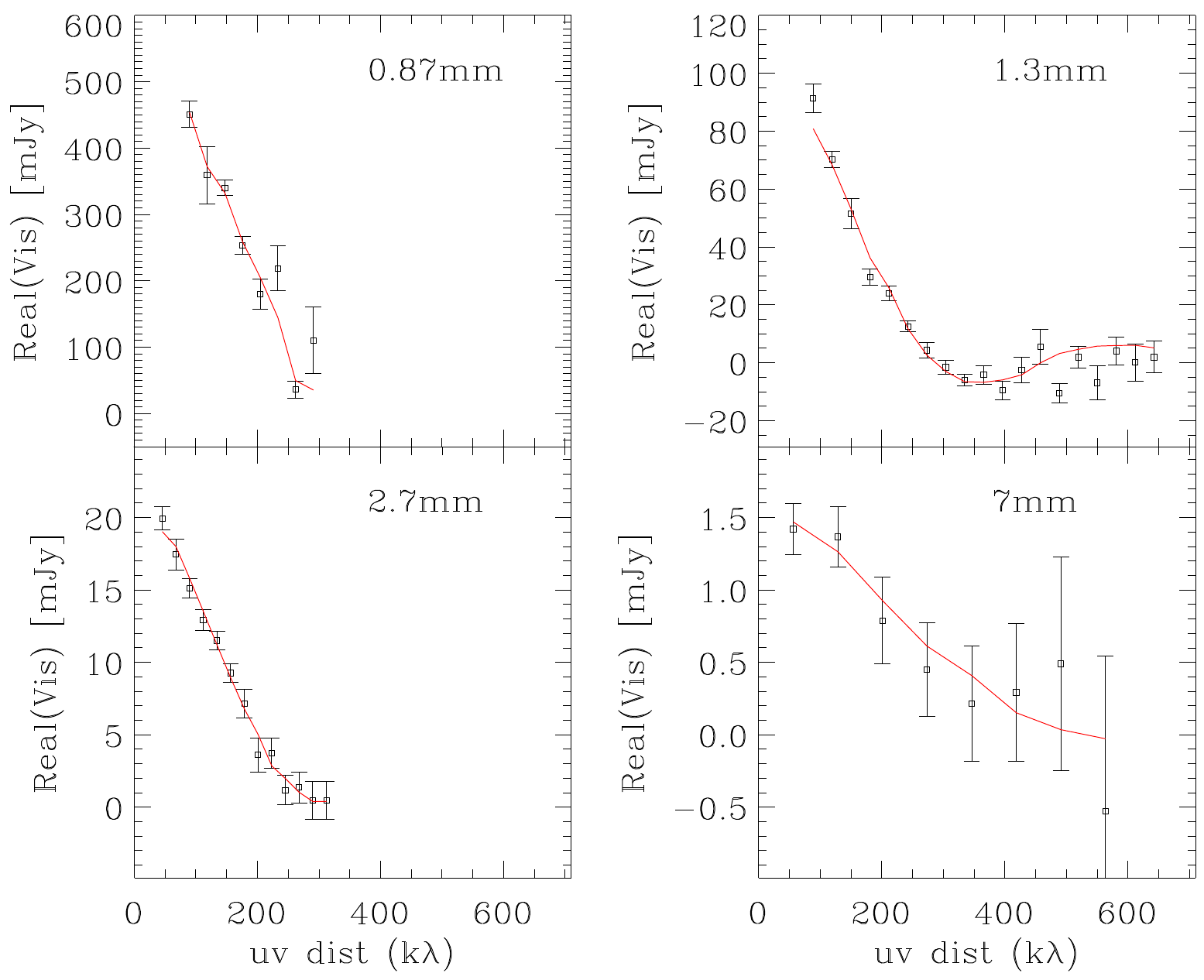}}
\caption{Deprojected radial averages of the real part of the visibility as a function of baseline length. The full lines 
show the predicted visibilities for the best fit models at each wavelength. 
}
\label{fuvfit} 
\end{figure}

In Fig.~\ref{fchi2} we show the projections of the $\chi^2$ hypercube on a few planes for each 
wavelength. 
The comparison between the observed images, the images
reconstructed from the visibilities of the best fitting models, and the residual image 
obtained from the different between the observed and the best fitting visibilities are
shown in Fig.~\ref{fmodima}. The comparison between the observed and best fitting real part
of the visibilities is shown directly in Fig.~\ref{fuvfit}, to produce the figure we deprojected the baseline 
lengths using the position angle and inclination of the disk as derived from the best fit, then we 
produced annular average visibilities for the data (points with error bars) and the model. 
In Table~\ref{tuvfit} we list the best fit values and the confidence range for the model 
parameters at each wavelength.

\begin{table}
\caption{Results of the model fits to the visibilities. For each parameter we 
report the best fit value and the minimum and maximum values of the
68\%\ confidence level intervals (1$\sigma$).}
\label{tuvfit}
\begin{tabular}{l|cccc}
\hline
Parameter&0.87~mm&1.3~mm&2.7~mm&7~mm\\
\hline
\\
$\Sigma_{\rm 40AU}$ (g\,cm$^{-2}$)&    7.1$^{+2.9}_{-1.5}$   &  2.5$^{+0.2}_{-0.1}$      &    2.5$^{+0.3}_{-0.3}$     &   5.0$^{+3.9}_{-3.0}$    \\
\\
p&  0.04$^{+0.4}_{-0.7}$    &      -0.075$^{+0.025}_{-0.125}$    &       0.5$^{+0.3}_{-0.25}$  &  0.5$^{+0.5}_{...}$   \\
\\
R$_{\rm out}$ (AU)& 54$^{+10}_{-5}$   &    63$^{+1}_{-2}$      &   70$^{+10}_{-6}$       &   52$^{+33}_{-12}$  \\
\\
\hline
\end{tabular}
\end{table}

The 1.3~mm dataset is the one with the best combination of resolution and signal to noise
and is the one for which we derive a more constrained fit for the disk model parameters. The best fit
values for the inclination and the position angle (measured east of north) are 34$^\circ$ and 42$^\circ$ respectively. 
The inclination is in good agreement with the value derived from observations of the CO emission \citep[note that these authors report the position angle of the disk axis]{2008A&A...488..565C}.
The slope of  the surface density distribution $p$ is constrained to be between -0.2 and 0,  implying an almost constant 
radial dust distribution. Values of $p$ close to 0 are also obtained from the observations at 0.87 mm, while $p \sim 0.5$ is 
suggested by the observations at longer wavelengths. As discussed in \citet{Iea2010}, a systematic variation of 
the fitted surface density distribution from 0.87 and 7 mm would suggest a radial variation of the slope of the dust 
opacity $\beta$. Values of  $p$ increasing with the wavelength would imply that $\beta$ increases with the distance 
from the central star \citep[from Eq. 9 in][]{Iea2010},  or that the maximum grain size decreases with the orbital distance. 
In particular, a variation of $p$ of about 0.5 between 0.87 and 7 mm,  corresponds to an increase of $\beta$ 
from 0 to about 1.4, between 0.2 and 63 AU. In terms of grain size distribution, this would imply
that the inner disk is mainly populated by dust grains of few centimeters in radius, while the outer disk is characterized by a sub-micron grain 
size distribution similar to what observed in the interstellar medium. 
Our results are in qualitative agreement with recent model predictions by \citet{Bea2010a} who predict
values of $\beta$ in circumstellar disks to grow as a function of radius due to the different growth 
level of grains as a function of radius as predicted by the dust evolution models of \citet{2008A&A...480..859B} and \citet{Bea2010b}.
Nevertheless, our initial results should be taken with caution: with the exception of the 1.3 mm observations, 
the angular resolution and the signal to noise level is 
too low for us to constrain reliably the value of $p$ at the various wavelengths, and more sensitive high 
angular resolution observations together with a proper fitting with a disk model which takes into account the 
dust variations as a function of radius are necessary to confirm our tentative conclusions.


From our observations, we measure an outer disk radius of about 60 AU, which is much smaller than the 
value of 140$\pm$14 AU derived from CO and dust continuum emission observations at 2\arcsec\ 
resolution \citep[note that we scaled the values of their Table 3 for a stellar distance of 100 pc]{2008A&A...488..565C}.   
When using a simple power-law description for the surface density, it is well known that 
disk radii mesured from the dust millimeter continuum emission are often smaller than those derived from 
molecular line observations \citep[e.g.][]{2008ApJ...678.1119H,2007A&A...469..213I}.
Consequently, the inferred radius of 60 AU  has to be considered as a lower limit.

\section{Summary and conclusions}

In this paper we have presented new high angular resolution observations from the submillimeter
through the centimeter wavelength range of the CQ~Tau system. At millimeter and submillimeter wavelegths
we detect and resolve the thermal dust emission from the circumstellar disk. At 3.6~cm we detect compact emission from the inner regions of the system that we attribute to gas emission. We fit the observations as 
the emission of a dusty disk with an additional contribution of unresolved gas emission.
The combination of high angular resolution observations
at several wavelengths and the accurate estimate of the possible contamination due to gas allows us
to derive more accurately the properties of the thermal dust emission from the circumstellar disk itself.

We find that the average value across the disk of the dust opacity coefficient $\beta$, computed between 1.3 and 7~mm,
is $\sim 0.6\pm0.1$. This confirms and puts on firmer ground previous derivations of $\beta$ in this system. These values of $\beta$ imply that significant grain growth has occurred in this disk, with the largest grains grown to $\sim$1~cm.
The exact value of the maximum grain size depends on the details of the dust model adopted and we find it likely to be in the range from 1~mm to 10~cm.

Our resolved images at several (sub)millimeter wavelengths show a weak indication of the signature of 
grain growth variation as a function of radius in the CQ~Tau disk. 
We thus find 
marginal evidence to support the expectation that the inner regions of the disk contain grains grown to 
larger sizes than the outer regions. This result needs to be confirmed by higher signal to noise data at
the extremes of the wavelength range probed by our observations.
In particular we do expect that in the near future EVLA observations at 7~mm and 1.3~cm will allow
to confirm (or disprove) our suggestion.

\begin{acknowledgements}We thank Luca Ricci for his help in computing the dust 
opacities and for many useful discussions on grain growth in protoplanetary disks.
All the observations used in this paper were taken in service mode
by the observatories staff at SMA, IRAM and NRAO, their excellent support is gratefully
acknowledged. LT and AN were partly supported by the grant ASI COFIS I/016/07/0. 
This work was performed in part under contract with the Jet Propulsion Laboratory 
(JPL) funded by NASA through the Michelson Fellowship Program. JPL is 
managed for NASA by the California Institute of Technology.
This research has made use of the SIMBAD database,
operated at CDS, Strasbourg, France.
\end{acknowledgements}

\bibliographystyle{aa}
\bibliography{15206}

\end{document}